\documentstyle[sprocl]{article}

\def\Journal#1#2#3#4{{#1} {\bf #2}, #3 (#4)}

\def\NPB{{\em Nucl. Phys.} B}
\def\PLB{{\em Phys. Lett.}  B}
\def\PRL{\em Phys. Rev. Lett.}
\def\PRD{{\em Phys. Rev.} D}
\def\ZPC{{\em Z. Phys.} C}
\def\PAN{\em Phys. of Atom. Nucl.}
\def\FBS{\em Few-Body Systems}
\def\PPNP{\em Prog. Part. Nucl. Phys.}
\def\lbc{\Lambda^0_b\to\Lambda^+_c+e^-+\bar\nu_e}
\def\lcs{\Lambda^+_c\to\Lambda+e^++\nu_e}
\def\lbcw{\Lambda_b^0\to\Lambda_c^+[\to\Lambda\pi^+] +
W^-[\to\ell^-\bar\nu_\ell]}

\def\be{\begin{equation}}
\def\ee{\end{equation}}
\def\bea{\begin{eqnarray}}
\def\eea{\end{eqnarray}}
\begin{document}

\title{SEMILEPTONIC DECAYS OF BOTTOM AND CHARM BARYONS
WITHIN RELATIVISTIC QUARK MODEL}

\author{M.A. IVANOV and V.E. LYUBOVITSKIJ}

\address{Bogoliubov Laboratory of Theoretical Physics, JINR,\\
Dubna Moscow Region, 141980, Russia}


\maketitle\abstracts{The results for observables of semileptonic decays
of bottom and charm baryons (Isgur-Wise functions, decay rates,
distributions, asymmetry parameters) are given within relativistic
quark model.  A comparison with recent experimental data
(CLEO Collaboration) is made.}

\section{Introduction}

The last few years have brought rapid development of the physics of hadrons
composed of light quarks $q$ ($u$, $d$ and $s$) and heavy quarks $Q$
($c$ or $b$). Heavy quarks are those whose masses
satisfy the condition $m_Q\gg \Lambda_{QCD}$, where $\Lambda_{QCD}$
is the QCD scale parameter. Weak decays of heavy hadrons are a unique tool
for determining the elements of the Cabibbo-Kobayashi-Maskawa matrix, for
studying phenomena lying outside the scope of the standard model, and also
for studying the internal structure of hadrons.

 From the theoretical point of view, this lively interest in weak decays of
heavy hadrons is mainly due to the discovery of a new type of spin-flavor
symmetry in the world of heavy quarks (the Isgur-Wise symmetry)
\cite{iws1,iws2} and to the development of the Heavy Quark Effective Theory
(HQET) \cite{iws1}$^-$\cite{jgk} for studying of heavy-hadron weak decays.
The Isgur-Wise symmetry (IWS) is occurring in limit of infinite mass of
{\it b} and {\it c} quarks - the heavy quark limit (HQL). The HQET is a
perturbative computational scheme based on expansion of QCD-inspired
effective Lagrangian in terms of inverse powers of the heavy quark mass.

The consequences of the IWS for weak decay form factors of hadrons
containing a single heavy quark were worked out by Isgur and Wise
\cite{iws1,iws2}. It was founded that these form factors are expressed
through four universal functions (Isgur-Wise functions) and satisfied
to group relations. Unfortunately, possibility to calculate
$\omega$-dependence of IW-functions using Standard Model is absent.
IW-functions are very sensitive to the effects of QCD at large distances,
it cannot be calculated in perturbation theory. Only normalizations of these
form factors at point $\omega=1$ are known. So calculation of IW-functions
got dissemination within various phenomenological approaches: QCD Sum Rules,
QCD on Lattice, Infinite Momentum Frame models, Quark Confinement Model,
Bag models and etc.

Our purpose is to describe observables of semileptonic decays of bottom
and charm baryons in the HQL within relativistic quark model.
This approach allows to take account long-distance effects of QCD
interactions. We shall calculate baryonic Isgur-Wise functions and charge
radii, decay rates and asymmetry parameters. Also detailed description of
$\lcs$ decay which was measured recently by CLEO Collaboration \cite{CLEO}
will be given.

\section{Model}

In our model \cite{aikl} baryons are considered as bound states of
valence quarks. Annihilation of baryons into quarks and {\it vice versa}
are described by means of the corresponding interaction Lagrangian.
\begin{eqnarray}
{\cal L}_{int}(x)=g_B  \bar B(x)J_B(x) + h.c.
\end{eqnarray}
\noindent where $B$ is a baryon field, $J_B$ is a three-quark current,
$g_B$ is a coupling constant.

\noindent A distribution of quarks in baryon is taken account by
corresponding relativistic vertex form factor
$F\biggl(\sum\limits_{i<j}[y_i-y_j]^2\biggr),$
where $y_i (i=1,2,3)$ are the space coordinates of quarks.
The three-quark current $J_B(x)$ is chosen in the form
\begin{eqnarray}
J_B(x)&=&\int dy_1\int dy_2 \int dy_3\,\, \delta
\biggl(x-\frac{\sum\limits_im_iy_i}{\sum\limits_im_i}\biggr)
\,\,F\biggl(\sum\limits_{i<j}[y_i-y_j]^2\biggr)\nonumber\\
& &\nonumber\\
&\times&q^{a_1}(y_1)q^{a_2}(y_2)q^{a_3}(y_3)\varepsilon^{a_1a_2a_3}
\end{eqnarray}
where the center of mass frame is used. Here spin and flavor indices are
omitted. It is easy to see that in the limit
\begin{eqnarray}
F\biggl(\sum\limits_{i<j}[y_i-y_j]^2\biggr)\to
\prod\limits_{i<j} \delta(y_i-y_j)\nonumber
\end{eqnarray}
\noindent the baryonic current $J_B(x)$ goes to
\begin{eqnarray}
J_B(x)=q^{a_1}(x)q^{a_2}(x)q^{a_3}(x)\varepsilon^{a_1a_2a_3}
\nonumber
\end{eqnarray}
\noindent To say other words the interaction of quarks becomes local.
The coordinates of quarks
are connected with Jacobi coordinates $\xi_1$ and $\xi_2$
by standard manner

\[\left\{
\begin{array}{l}
y_1=x-3\xi_1(m_2+m_3)/\sum\limits_i m_i\\[3mm]
y_2=x+3\xi_1{m_1}/\sum\limits_i m_i -
2\sqrt{3}\xi_2m_3/(m_2+m_3)\\[3mm]
y_3=x+3\xi_1{m_1}/\sum\limits_i m_i +
2\sqrt{3}\xi_2m_2/(m_2+m_3)\\
\end{array}
\right.  \]

For light baryons we apply the SU(3)-symmetric picture based on unitary
symmetry in the light quark sector. For heavy-light baryons the heavy quark
limit is used ($m_Q\gg m_q$, $m_Q\to\infty$). Hence heavy quark removes
to the c.m. of heavy-light baryon. Therefore coordinates of quarks in
heavy-light and light baryons look like this

\vspace*{0.5cm}
\unitlength=1mm
\special{em:linewidth 0.4pt}
\linethickness{0.4pt}
\begin{picture}(100.00,30.00)
\put(20.00,23.00){\circle{14.00}}
\put(20.00,23.00){\circle*{4.00}}
\put(24.00,26.00){\circle*{2.00}}
\put(24.00,20.00){\circle*{2.00}}
\put(84.00,23.00){\circle{14.00}}
\put(81.00,23.00){\circle*{2.00}}
\put(86.00,26.00){\circle*{2.00}}
\put(86.00,20.00){\circle*{2.00}}
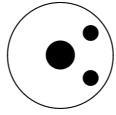
\put(21.00,8.00){\makebox(0,0)[cc]{Heavy-Light Baryon}}
\put(84.00,8.00){\makebox(0,0)[cc]{Light Baryon}}
\end{picture}

\hspace*{1.8cm}{\Large $\Downarrow$  \hspace*{5.85cm}    $\Downarrow$}
\begin{eqnarray}
\left\{
\begin{array}{c}
y_Q=y_1=x\\[3mm]
y_{q_1}=y_2=x+3\xi_1 - \sqrt{3}\xi_2\\[3mm]
y_{q_2}=y_3=x+3\xi_1 + \sqrt{3}\xi_2\\[3mm]
\sum\limits_{i<j}[y_i-y_j]^2=18(\xi_1^2+\xi_2^2)
\end{array}\right.
%\hspace*{1.75cm}
\hspace*{1.5cm}
\left\{
\begin{array}{c}
y_{q_1}=y_1=x-2\xi_1\\[3mm]
y_{q_2}=y_2=x+\xi_1 - \sqrt{3}\xi_2\\[3mm]
y_{q_3}=y_3=x+\xi_1 + \sqrt{3}\xi_2\\[3mm]
\sum\limits_{i<j}[y_i-y_j]^2=18(\xi_1^2+\xi_2^2)
\end{array}\right.
\nonumber
\end{eqnarray}

\vspace*{0.5cm}
The form of vertex function $F\biggl(\sum\limits_{i<j}[y_i-y_j]^2\biggr)$
allows us to make all matrix elements ultroviolet finite. One has to
underline that vertex functions can be understood as phenomenological
taking account of long-distance effects of QCD interactions. In this point
we follow the ideas of QCD-inspired models of hadrons
\cite{cdr}$^-$\cite{yuk} based on bilocal procedure of bosonization.

For light baryons with spin-parity $J^P={\frac{1}{2}}^+$
two independent forms of interaction Lagrangians exist (so-called
{\it vector variant} and {\it tensor variant}) \cite{fbs}
\begin{eqnarray}
L^{light}_{int}(x)&=&g_{B_q}\bar B^{mn}(x)
\int d\xi_1\int d\xi_2 \,\,f(\xi_1^2+\xi_2^2)
\Gamma_1 \lambda_f^{nk_1}q^a_{k_1}(x-2\xi_1)\nonumber \\
& &\nonumber \\
&\times&q^b_{k_2}(x+\xi_1-\sqrt{3}\xi_2)
C\Gamma_2 \lambda_f^{kk_3}
q^c_{k_3}(x+\xi_1+\sqrt{3}\xi_2)\nonumber
\varepsilon^{abc}\varepsilon^{mk_2k}+h.c.\nonumber
\end{eqnarray}

\vspace*{0.3cm}
\[B=\left(
\begin{array}{ccc}
{1\over\sqrt2}\Sigma^0+{1\over\sqrt6}\Lambda^0 & \Sigma^+
 & p \\
\Sigma^- & -{1\over\sqrt2}\Sigma^0+{1\over\sqrt6}\Lambda^0
& n \\
-\Xi^- & \Xi^0 & -{2\over\sqrt6}\Lambda^0 \\
\end{array}
\right) \mbox{baryon matrix} \]

\[q_j^a=\left(
\begin{array}{c}
u^a \\
d^a \\
s^a \\
\end{array}
\right)  \mbox{set of quark fields} \]

\vspace*{0.2cm}
\[\Gamma_1\otimes C\Gamma_2=\left\{
\begin{array}{ll}
\gamma^\mu\gamma^5\otimes C\gamma_\mu
&\,\,\,\,\mbox{vector variant}\\[3mm]
\sigma^{\mu\nu}\gamma^5\otimes C\sigma_{\mu\nu}
&\,\,\,\,\mbox{tensor variant}
\end{array}\right. \]
\noindent One has to remark that spin-flavor structure of our light
baryon currents is identical to ones used in QCD Sum Rules by
Ioffe et al. \cite{iof}

Choice of quark currents of heavy-light baryons was discussed firstly
by Shuryak in ref.\cite{shur}. He has showed that for $\Lambda$-type baryons
($\Lambda_Q$ or $\Xi_Q$) containing a scalar light diquark
three possibilities of baryonic currents exist
\[\left\{
\begin{array}{ll}
J^{(1)}_{\Lambda_Q}=Q^a(u^bC\gamma^5d^c)\varepsilon^{abc}
&\mbox{scalar variant}\\[3mm]
J^{(2)}_{\Lambda_Q}=\gamma_5Q^a(u^bCd^c)\varepsilon^{abc}
&\mbox{pseudoscalar variant}\\[3mm]
J^{(3)}_{\Lambda_Q}=\gamma_\mu Q^a(u^bC\gamma^\mu\gamma^5d^c)
\varepsilon^{abc}
&\mbox{axial variant}
\end{array}\right. \]
and for $\Omega$-type baryons
($\Omega_Q$, $\Omega_Q^\star$ or $\Sigma_Q$, $\Sigma_Q^\star$) containing
a vector light diquark two possibilities exist
\vspace*{0.3cm}
\[\left\{
\begin{array}{ll}
J^{(1)}_{\Omega_Q}=\gamma_\mu\gamma_5Q^a(s^bC\gamma^\mu s^c)
\varepsilon^{abc}&\mbox{vector variant}\\[3mm]
J^{(2)}_{\Omega_Q}=\sigma_{\mu\nu}\gamma_5Q^a
(s^bC\sigma^{\mu\nu}s^c)\varepsilon^{abc}&\mbox{tensor variant}
\end{array}\right. \]
\vspace*{0.3cm}
\[\left\{
\begin{array}{ll}
J^{\mu(1)}_{\Omega^\star_Q}=Q^a(s^bC\gamma^\mu s^c)\varepsilon^{abc}
&\mbox{vector variant}\\[3mm]
J^{\mu(2)}_{\Omega^\star_Q}=-i\gamma_\nu Q^a
(s^bC\sigma^{\mu\nu}s^c)\varepsilon^{abc}
&\mbox{tensor variant}
\end{array}\right. \]
In our calculations we will use {\it tensor variant} for light pseudoscalar
baryons, {\it scalar variant} for $\Lambda$-type baryons and
{\it vector variant} for $\Omega$-type baryons. One has to remark
that {\it pseudoscalar variant} for $\Lambda$-type baryons doesn't
give contribution at the HQL.

Matrix elements of semileptonic decays of heavy-light baryons are
described in our model by triangle diagram (see, Fig.1). As the light
quark propagator the standard free fermion propagator is used
\begin{eqnarray}
S_q(k)=\frac{i(\not\! k+m_q)}{k^2-m_q^2+i\epsilon}
\end{eqnarray}
\noindent where $m_q$ is the light quark mass which is a free parameter.
\noindent As the heavy quark propagator we use propagator arising
at the heavy quark limit (HQL)
\begin{eqnarray}
S(k+v\bar\Lambda)=\frac{i(1+\not\! v)}{2(v\cdot k+\bar\Lambda+i\epsilon)}
\end{eqnarray}
\noindent where $\bar\Lambda$ is the difference between masses of heavy
baryon and heavy quark at the HQL, $v$ is the four-velocity of heavy baryon.

\begin{center}
\unitlength=.9mm
\special{em:linewidth 0.4pt}
\linethickness{0.4pt}
\begin{picture}(95.00,68.00)
\put(50.00,13.00){\oval(40.00,16.00)[]}
\put(70.00,13.00){\circle*{2.00}}
\put(30.00,13.00){\circle*{2.00}}
\put(25.00,13.00){\circle*{4.00}}
\put(75.00,13.00){\circle*{4.00}}
\put(75.00,13.00){\line(1,0){20.00}}
\put(75.00,14.00){\line(1,0){20.00}}
\put(75.00,12.00){\line(1,0){20.00}}
\put(25.00,12.00){\line(-1,0){20.00}}
\put(25.00,13.00){\line(-1,0){20.00}}
\put(25.00,14.00){\line(-1,0){20.00}}
\put(25.50,13.90){\line(5,6){24.00}}
\put(25.00,13.00){\line(5,6){25.00}}
\put(24.50,12.10){\line(5,6){24.00}}
\put(49.90,42.90){\line(5,-6){24.00}}
\put(50.00,43.00){\line(5,-6){25.00}}
\put(50.10,43.10){\line(5,-6){24.00}}
\put(50.00,43.00){\line(0,1){3.00}}
\put(50.00,47.00){\line(0,1){3.00}}
\put(50.00,51.00){\line(0,1){3.00}}
\put(50.00,55.00){\line(0,1){3.00}}
\put(50.00,59.00){\line(0,1){3.00}}
\put(55.00,48.00){\line(0,1){12.00}}
\put(69.00,53.00){\makebox(0,0)[cc]{$q=p-p^\prime$}}
\put(50.00,65.00){\makebox(0,0)[cc]{$\ell\nu_\ell$}}
\put(0.00,13.00){\makebox(0,0)[cc]{$B^\prime$}}
\put(100.00,13.00){\makebox(0,0)[cc]{$B$}}
\put(88.00,19.00){\makebox(0,0)[cc]{$p$}}
\put(12.00,19.00){\makebox(0,0)[cc]{$p^\prime$}}
\put(33.00,33.00){\makebox(0,0)[cc]{$p^\prime+k$}}
\put(68.00,33.00){\makebox(0,0)[cc]{$p+k$}}
\put(50.00,26.00){\makebox(0,0)[cc]{${k^\prime-k\over 2}$}}
\put(50.00,0.00){\makebox(0,0)[cc]{${k^\prime+k\over 2}$}}
\put(12.00,13.00){\line(2,1){4.00}}
\put(12.00,13.00){\line(2,-1){4.00}}
\put(88.00,13.00){\line(2,1){4.00}}
\put(88.00,13.00){\line(2,-1){4.00}}
\put(55.00,60.00){\line(-1,-4){1.00}}
\put(55.00,60.00){\line(1,-4){1.00}}
\put(48.00,21.00){\line(4,1){4.00}}
\put(48.00,21.00){\line(4,-1){4.00}}
\put(52.00,5.00){\line(-4,1){4.00}}
\put(52.00,5.00){\line(-4,-1){4.00}}
\put(38.00,29.00){\line(1,2){2.00}}
\put(38.00,29.00){\line(2,1){4.00}}
\put(59.00,32.00){\line(2,-1){4.00}}
\put(59.00,32.00){\line(1,-4){1.00}}
\put(34.00,13.00){\makebox(0,0)[cc]{$\Gamma^\prime_1$}}
\put(66.00,13.00){\makebox(0,0)[cc]{$\Gamma^\prime_2$}}
\put(25.00,8.00){\makebox(0,0)[cc]{$\Gamma_1$}}
\put(75.00,8.00){\makebox(0,0)[cc]{$\Gamma_2$}}
\put(45.00,43.50){\makebox(0,0)[cc]{$O_\mu$}}
\end{picture}
\end{center}

\vspace*{0.3cm}
\begin{center}
Fig. 1
\end{center}

\vspace*{0.5cm}
Typical matrix element describing heavy-to-heavy
transition is given by the following expression

\begin{eqnarray}
\bar u(v^\prime)M_\Gamma(v,v^\prime)u(v)&=&
gg^\prime\hspace*{-0.1cm}\int\frac{d^4k}{\pi^2i}
\int\frac{d^4k^\prime}{\pi^2i}\hspace*{0.1cm}
{\rm Tr}\bigg[\Gamma_1^\prime S_q\biggl(\frac{k^\prime-k}{2}\biggr)
\Gamma_2^\prime S_q\biggl(\frac{k^\prime+k}{2}\biggr)\biggr]\nonumber \\
& &\nonumber \\
&\times&F^2(9k^2+3k^{\prime 2})
\bar u(v^\prime)\Gamma_1 S(k+v^\prime\bar\Lambda)
\Gamma S(k+v\bar\Lambda)\Gamma_2u(v)
\nonumber
\end{eqnarray}
\noindent Here coupling constants $g$ and $g^\prime$ are calculated by
solving the equation:
\begin{eqnarray}
Z_{B}=1-g^2\Pi_B^\prime(M_{B})=0
\nonumber
\end{eqnarray}
\noindent It is the so-called compositness condition in quantum field
theory, which coincides with the Ward identity between the derivative of
baryon mass operator $\Pi_B^\prime$ and electromagnetic vertex function.
It provides the correct normalization of baryonic IW-functions.

Now let us to discuss the choice of free parameters in our approach.
There are three groups of free parameters in our model:
light quark masses, the set of parameters $\bar\Lambda$
and vertex  functions.

For masses of $u$ and $d$ quarks the unit parameter is used:
$m_u=m_d=m_q$, which is varied in the limits 310-340 MeV. The best fit
$m_q$=315 MeV comes from analysis of nucleon physics within our model.
The strange quark mass $m_s$ is varied in the limits 500-550 MeV.
The best value for $m_s$=500 MeV comes from analysis of
$\Lambda_c^+\to\Lambda\ell^+\nu_\ell$ decay.
The values of parameters $\bar\Lambda$ must depend on the flavor of light
diquark. We suggest that $\bar\Lambda$ must be little than the sum
of light quark masses: $\bar\Lambda< m_{q_1}+m_{q_2}$. This constraint
provides the absence of singularities in the matrix elements connected with
production of free quarks. Finally, we use $\bar\Lambda=600$ MeV for
heavy-light baryon without strange quark, $\bar\Lambda=800$ MeV for
heavy-light baryon containing a single strange quark and
$\bar\Lambda=950$ MeV for heavy-light baryon containing two strange quarks.
For simplicity, the Gaussian form of vertex functions is used
$$F(\xi_1^2+\xi_2^2)=\frac{\Lambda^4}{(16\pi^2)}
\exp\biggl([\xi_1^2+\xi_2^2]\frac{\Lambda^2}{4}\biggr)$$
\noindent where $\Lambda$ is a cutoff parameter. The parameters
$\Lambda$ must be different for light and heavy-light (h.-l.) baryons.
The value of $\Lambda$ for light baryons was fixed in the analysis
of nucleon properties: $\Lambda_{light}$=2.885 GeV. The value of
$\Lambda_{h.-l.}=1.909$ GeV was found in the analysis of
$\Lambda_c^+\to\Lambda\ell^+\nu_\ell$ process.

\section{Results}

In this Section we focus on the results of our calculations. We present the
predictions for the $b\to c$ semileptonic decays. IW-functions, decay rates
and asymmetry parameters in two-cascade decay $\lbcw$ are calculated. Also
heavy-to-light flavor exchange decays are considered. The detailed
description of the $\lcs$ decay which was recently measured by CLEO
Collaboration \cite{CLEO}is given. Here the following values for CKM matrix
elements are used:
$$|V_{bc}|=0.044, |V_{cs}|=1, |V_{cd}|=0.204, |V_{bu}|=0.002\div 0.005$$

\subsection{Isgur-Wise functions}

It is well-known, that weak baryonic currents for $b\to c$
transitions are expressed through the three universal Isgur-Wise functions
$\zeta, \xi_1, \xi_2$ at the HQL.

\vspace*{0.5cm}
$\bullet$  $\Lambda_b\to\Lambda_c$ Transition

\begin{eqnarray}
<\Lambda_c(v^\prime)|\bar c \,\Gamma \,b|\Lambda_b(v)>=
\zeta(\omega)\bar u_{\Lambda_c}(v^\prime)\Gamma u_{\Lambda_b}(v)
\nonumber
\end{eqnarray}

\vspace*{0.5cm}
$\bullet$ $\Omega_b\to\Omega_c(\Omega_c^\star)$ Transition

\begin{eqnarray}
& &<\Omega_c(v^\prime)\,\mbox{or}\,\Omega_c^\star(v^\prime)|
\bar c \,\Gamma \, b|\Omega_b(v)>=
\bar B^\mu_c(v^\prime) \,\Gamma \,B^\nu_b(v)[-\xi_1(\omega)g_{\mu\nu}
+\xi_2(\omega)v_\mu v^\prime_\nu]\nonumber \\
& &\nonumber \\
& &B^\mu_Q(v)=\frac{\gamma^\mu+v^\mu}{\sqrt{3}}u_{\Omega_Q}(v),\,\,
B^\mu_Q(v)=u^\mu_{\Omega^\star_Q}(v)
\nonumber
\end{eqnarray}

In our model IW-functions are expressed through the three structure
integrals $\Phi_i, i=0,1,2$
\begin{eqnarray}
& &\zeta(\omega)=\frac{\Phi_0(\omega)}{\Phi_0(1)}, \;\;
\xi_1(\omega)=\frac{\Phi_1(\omega)}{\Phi_1(1)}, \;\;
\xi_2(\omega)=\frac{\Phi_2(\omega)}{\Phi_1(1)}\nonumber  \\
& &\nonumber\\
& &\Phi_{\rm I}(\omega)=
\int\limits_0^\infty dxx\int\limits_0^\infty \frac{dyy}{(y+1)^2}
\int\limits_0^1 d\phi\int\limits_0^1 d\theta \,\, R_I(\omega) \,\,
\exp\biggl[-4s\biggl(\mu_q^2-\frac{\bar\lambda^2}{4}\biggr)\biggr]
\nonumber \\
& &\nonumber\\
&\times&\exp\biggl[-2x^2s\phi(1-\phi)(\omega-1)-
s(x-\bar\lambda)^2-\frac{4}{3}\mu_q^2(1-2\theta)^2\frac{y^2}{1+y}\biggr]
\nonumber
\end{eqnarray}
\noindent where

\begin{eqnarray}
R_0(\omega)&=&\mu_q^2+\frac{1}{s(1+y)}+
\frac{x^2\beta}{4(1+y)^2}(1+2\phi(1-\phi)(\omega-1))\nonumber \\
& &\nonumber\\
R_1(\omega)&=&\mu_q^2+\frac{1}{2s(1+y)}+
\frac{x^2\beta}{4(1+y)^2}(1+2\phi(1-\phi)(\omega-1))\nonumber \\
& &\nonumber\\
R_2(\omega)&=&\frac{x^2\beta}{2(1+y)^2}
\phi(1-\phi)\nonumber \\
& &\nonumber\\
\beta&=&1+2y+4y^2\theta(1-\theta), \,\,\,\,\,\,
s=\frac{2}{3}+\frac{\beta}{3(1+y)},
\,\,\,\,\,\, \mu_q=\frac{m_q}{\Lambda},\,\,\,\,
\bar\lambda=\frac{\bar\Lambda}{\Lambda} \,\,\,\,\,
\nonumber
\end{eqnarray}
\noindent It is clear that functions $\zeta$ and $\xi_1$ have the
correct normalization at zero recoil: $\zeta(1)=1$ and $\xi_1(1)=1$.

Now let us to check the model-independent inequalities for form
factors of $\Omega_b$ decays derived by Xu \cite{xu}
\begin{eqnarray}
1&\geq&\frac{2+\omega^2}{3}\xi_1^2(\omega)+
\frac{(\omega^2-1)^2}{3}\xi_2^2(\omega)
+\frac{2}{3}(\omega-\omega^3)\xi_1(\omega)\xi_2(\omega)\\
& &\nonumber \\
\rho^2_{\xi_1}&\ge&\frac{1}{3}-\frac{2}{3}\xi_2(1)
\end{eqnarray}

Exploiting the expression for $\xi_1(\omega)$ and $\xi_2(\omega)$
functions we can proof that the inequality (6)
for the slope of $\xi_1$ function is fulfilled for any values of
$\omega$. Also we obtain the low limit for the radius of $\xi_1$ function
\begin{eqnarray}
\rho^2_{\xi_1}\geq 1/3
\nonumber
\end{eqnarray}
\noindent Additionally we find that function $\xi_2(\omega)$ at point
$\omega=1$ satisfies the condition $0<\xi_2(1)<1/2$.

There is more sophisticated situation with the inequality (5). For
convenience, we rewrite this inequality in the form
\begin{eqnarray}
1\geq B(\omega)=\frac{1}{3}\xi_1^2(\omega)+
\frac{2}{3}(\omega\xi_1(\omega)-\xi_2(\omega)(\omega^2-1))^2
\nonumber
\end{eqnarray}
\noindent We can show that the combination $\omega\xi_1(\omega)-
\xi_2(\omega)(\omega^2-1)$ satisfies to the following condition
$0<\omega\xi_1(\omega)-\xi_2(\omega)(\omega^2-1)\leq\omega\xi_1(\omega)$.
Hence,
\begin{eqnarray}
\frac{1}{3}\xi_1^2(\omega)< B(\omega)
\leq \frac{1+2\omega^2}{3}\xi_1^2(\omega)
\nonumber
\end{eqnarray}

Thus the Bjorken-Xu inequality (5) gives us the upper limit for the function
$\xi_1(\omega)$:
\begin{eqnarray}
\xi_1\leq\sqrt{\frac{3}{2\omega^2+1}}
\nonumber
\end{eqnarray}
\noindent Of course, free parameters of the model were chosen with taking
into account of the last constraint.

On the Fig.2 the results for the IW-function $\zeta$ are given.
One can see, our result coincides with dipole fit result \cite{desy} and
goes higher than results of IMF models \cite{kroll,desy}.

Also we calculate the radii of IW-functions $\zeta$ and $\xi_1$
which are defined as

\begin{eqnarray}
\rho^2_F=-F^\prime(1),
\hspace*{0.5cm}F^\prime(\omega)=\frac{dF(\omega)}{d\omega}
\hspace*{1cm} \mbox{where} \hspace*{0.5cm}F=\zeta \hspace*{0.5cm}
\mbox{or}\hspace*{0.5cm} \xi_1\nonumber
\end{eqnarray}
The results for the charge radii are listed in Table 1.
For comparison we remember the results for these quantities predicted
by IMF model \cite{desy} and dipole model \cite{desy}.

\vspace*{1cm}
\begin{center}
Table 1. Radii of Isgur-Wise Functions
\end{center}
\def\arraystretch{2.5}
\begin{center}
\begin{tabular}{|c|c|c|c|c|c|c|}
\hline\hline
Model & $\rho^2_{\zeta}$ & $\rho^2_{\xi_1}$ \\
\hline\hline
Our  & 1.70 & 1.74 \\
\hline
K\"{o}rner et al. \cite{desy}& 3.04 & - \\
\hline
Dipole Fit \cite{desy}& 1.78 & - \\
\hline\hline
\end{tabular}
\end{center}

\newpage
\subsection{Decay Rates and Asymmetry Parameters}

In the Table 2 the results for total and partial rates of
$\lbc$ decay are given

\vspace*{.7cm}
\begin{center}
Table 2. Rates of $\lbc$ Decay (in units $10^{10}$ sec$^{-1}$)
\end{center}
\def\arraystretch{2.5}
\begin{center}
\begin{tabular}{|c|c|c|c|c|c|c|c|c|}
\hline\hline
Approach & $\Gamma_{total}$ & $\Gamma_T$ & $\Gamma_{T_+}$ & $\Gamma_{T_-}$ &
$\Gamma_L$ & $\Gamma_{L_+}$ & $\Gamma_{L_-}$ \\
\hline\hline
Our & 5.66 & 2.19 & 0.56 & 1.63 & 3.47 & 0.13 & 3.34 \\
\hline
K\"{o}rner et al. \cite{desy} & 3.71 & 1.58 & 0.43 & 1.15 & 3.13 & 0.10 & 2.03
\\
\hline
Dipole Fit \cite{desy} & 5.65 & 2.17 & 0.56 & 1.61 & 3.48 & 0.13 & 3.35 \\
\hline\hline
\end{tabular}
\end{center}

\vspace*{.7cm}
In Table 3 we give the predictions for the asymmetry parameters which
characterize the two-cascade decay $\lbcw$

\vspace*{.7cm}
\begin{center}
Table 3. Asymmetry Parameters of
$\Lambda_b^0\to\Lambda_c^+[\to\Lambda\pi^+] + W^-[\to\ell^-\bar\nu_\ell]$
Decay.
\end{center}
\def\arraystretch{2.5}
\begin{center}
\begin{tabular}{|c|c|c|c|c|c|c|}
\hline\hline
Model & $\alpha$ & $\alpha^\prime$ & $\alpha^{\prime\prime}$ &
$\gamma$ & $\alpha_P$ & $\gamma_P$ \\
\hline\hline
Our  & -0.76 & -0.12 & -0.52 & 0.56 & 0.38 & -0.17\\
\hline
K\"{o}rner et al. \cite{desy}& -0.71 & -0.12 & -0.46 & 0.61 & 0.33 & -0.19\\
\hline
Dipole Fit \cite{desy}& -0.75 & -0.12 & -0.51 & 0.57 & 0.37 & -0.17\\
\hline\hline
\end{tabular}
\end{center}

\vspace*{.7cm}
The results for rates of various modes of semileptonic decays of bottom
and charm baryons are performed in Table 4.

\newpage
\def\arraystretch{2.5}
\begin{center}
Table 4. Semileptonic Decay Rates (in units 10$^{10}$ s$^{-1}$).
\end{center}
\begin{center}
\begin{tabular}{|c|c|c|c|c|c|} \hline
 process & Ref.\cite{sin} & Ref.\cite{cheng} & Ref.\cite{kor}
& Our & Exp. \cite{PDG} \\
\hline
$\Lambda_c^+\to\Lambda^0 e^+\nu_e$  & 9.8 & 7.1 &
& 6.21 & 7.0$\pm$ 2.5 \\
\hline
$\Xi_c^0\to\Xi^- e^+\nu_e$  & 8.5 & 7.4 & & 4.66 & \\
\hline
$\Lambda_b^0\to p e^-\nu_e$ & & & & 0.006$\div$ 0.034 &\\
\hline
$\Lambda_c^+\to ne^+\nu_e$  & & & &  1.04 & \\
\hline
$\Lambda_b^0\to\Lambda_c^+ e^-\bar{\nu}_e$  & 5.9 & 5.1 & 5.14 & 5.66
& \\
\hline
$\Xi_b^0\to\Xi_c^+ e^-\bar{\nu}_e$  & 7.2 & 5.3 & 5.21 & 4.67
& \\
\hline
$\Omega_b^-\to\Omega_c^0 e^-\bar{\nu}_e$
& 5.4 & 2.3 & 1.52 & 0.77 & \\
\hline
$\Omega_b^-\to\Omega_c^{\star 0} e^-\bar{\nu}_e$
& & & 3.41 & 2.02 & \\
\hline
$\Sigma_b^+\to\Sigma_c^{++} e^-\bar{\nu}_e$
& 4.3 & & & 1.42 & \\
\hline
$\Sigma_b^{+}\to\Sigma_c^{\star ++} e^-\bar{\nu}_e$
& & & & 3.26 & \\
\hline
\end{tabular}
\end{center}

\vspace*{0.5cm}
\subsection{Decay $\lcs$}

In this section we focus on the properties of the $\lcs$ decay which was
recently investigated by CLEO Collaboration \cite{CLEO}. At the HQL, when
a mass of charm quark goes to infinity ($m_C\to\infty$), weak hadronic
current is defined by two form factors $f_1$ and $f_2$ as

\begin{eqnarray}
<\Lambda(p^\prime)|\bar sO_\mu c|\Lambda_c(v)>=
\bar u_\Lambda(p^\prime)[f_1(p^\prime\cdot v)+\not \! v
f_2(p^\prime\cdot v)]O_\mu u_{\Lambda_c}(v)
\nonumber
\end{eqnarray}

In order to extract the form factor ratio $R=f_2/f_1$
>from the CLEO experiment \cite{CLEO} an assumption was made about
the $q^2$ dependence of the form factors $f_1$ and $f_2$.
Following the model of K\"{o}rner and Kr\"{a}mer \cite{kk},
the identical dipole form of weak form factors was used. The same
{\it ansatz} was used in the ref. \cite{cheng}. In our model, the form
factors $f_1$ and $f_2$ have the different $q^2$ dependence. In Table 5 we
give the results for ratio $R$ obtained within our approach at maximum
and zero recoils. You can see, that our predictions weakly deviate from
experimental data \cite{CLEO} and result of Cheng and Tseng \cite{cheng}.
The momentum dependence of form factors ($f_1$ and $f_2$) and their ratio
is drawn on Fig.3. Here we use the notation variable
$\omega=(p^\prime\cdot v)/M_\Lambda$, where $M_\Lambda$ is the $\Lambda$
baryon mass. We found that the variation of the value of $R$ is small.
Also, in Table 5, our results for rate of $\lcs$ transition and for
asymmetry parameter $\alpha_{\Lambda_C}$ are given. They are in a good
agreement with experimental data and result of model \cite{cheng}.

\vspace*{0.5cm}
\begin{center}
Table 5. Properties of $\lcs$ Decay
\end{center}
\begin{center}
\def\arraystretch{2.0}
\begin{tabular}{|c|c|c|c|}\hline
Value  & Our & Ref. \cite{cheng}& Exp. \cite{CLEO,PDG}\\
\hline
& & & \\
$\Gamma(\Lambda_c^+\to\Lambda^0)$ & 6.21 & 7.1 & 7.0$\pm$ 2.5 \\
in 10$^{-10}$ s$^{-1}$ & & & \\
\hline
& & & \\
$\alpha_{\Lambda_c}$ & -0.84 & & -0.82$^{+0.09+0.06}_{-0.06-0.03}$\\
& & & \\
\hline
& & & \\
$R=f_2/f_1$ & -0.23 (q$^2$=q$^2_{max}$) & -0.23 (q$^2$=q$^2_{max}$)
& -0.25$\pm$ 0.14$\pm$ 0.08\\ & -0.18 (q$^2$=0) & & \\
\hline
\end{tabular}
\end{center}

\newpage
\section*{Acknowledgments}
One of us (V. L.) would like to thank Organizing Committee
of the conference "HADRON-95" for the financial support which have
made possible the participation at the conference.


\section*{References}
\begin{thebibliography}{99}

\bibitem{iws1}N. Isgur and M. Wise \Journal{\PLB}{232}{113}{1989};
B {\bf 237}, {527} {(1990)}.

\bibitem{iws2}N. Isgur and M. Wise \Journal{\NPB}{348}{276}{1991}.

\bibitem{geo}H. Georgi, \Journal{\PLB}{240}{447}{1990}.

\bibitem{gri}B. Grinstein, \Journal{\NPB}{339}{253}{1990}.

\bibitem{eic}E. Eichten \& B. Hill, \Journal{\PLB}{234}{511}{1990}.

\bibitem{fal}A.F. Falk, et al., \Journal{\NPB}{343}{1}{1990}.

\bibitem{man}T. Mannel, W. Roberts \& Z. Ryzak,
\Journal{\NPB}{355}{38}{1991}.

\bibitem{jgk}J.G. K\"{o}rner \& G. Thompson,
\Journal{\PLB}{264}{185}{1991}.

\bibitem{CLEO}T. Bergfeld, et al, {\em Preprint} {\bf CLEO 94-4} (1994);\\
%G. Crawford, et al, {\em Preprint} {\bf CLEO 94-24} (1994).
G. Crawford, et al, \Journal{\PRL}{75}{624}{1995}.

\bibitem{aikl}I. Anikin, M. Ivanov, N. Kulimanova and V. Lyubovitskij,\\
\Journal{\ZPC}{65}{681}{1995}; \Journal{\PAN}{57}{1082}{1994}.

\bibitem{cdr}C.D. Roberts \& R.T. Cahill, \Journal{\PRD}{32}{2419}{1985}.

\bibitem{gh}T. Goldman \& R.W. Haymaker, \Journal{\PRD}{24}{724}{1981}.

\bibitem{yuk}Yu. Kalinovsky et al., \Journal{\PLB}{231}{288}{1989}.

\bibitem{fbs}G.V. Efimov, M.A. Ivanov and V.E. Lyubovitskij,\\
\Journal{\FBS}{6}{17}{1989}.

\bibitem{iof}B.L. Ioffe, \Journal{\NPB}{188}{317}{1981}.

\bibitem{shur}E.V. Shuryak, \Journal{\NPB}{198}{83}{1982}.

\bibitem{xu}Q.P. Xu, \Journal{\PRD}{48}{5429}{1993}.

\bibitem{kroll}X.-H. Guo \& P. Kroll, \Journal{\ZPC}{59}{567}{1993}.

\bibitem{desy}B. K\"{o}nig, et al., {\em Preprint} {\bf DESY 93-011} (1993).

\bibitem{sin}R. Singleton, \Journal{\PRD}{43}{2939}{1991}.

\bibitem{cheng}H.-Y. Cheng \& B. Tseng, {\em Preprint} {\bf hep-ph/9502391}
(1995).

\bibitem{kor}J.G. K\"{o}rner, et al., \Journal{\PPNP}{33}{787}{1994}.

\bibitem{PDG}Particle Data Group, \Journal{\PRD}{50}{1173}{1994}.

\bibitem{kk}J.G. K\"{o}rner \& M. Kr\"{a}mer, \Journal{\PLB}{275}{495}{1992}.

\end{thebibliography}
\end{document}